\documentclass[prl,amsfonts,amsmath,showpacs,showkeys,preprint]{revtex4}
\usepackage{graphicx}
\usepackage{dcolumn}
\usepackage{bm}
\usepackage{eepic}
\bibpunct{[}{]}{,}{a}{}{;}
\RequirePackage{times}
\RequirePackage{mathptm}
\usepackage{longtable}

\begin{document}

\title{The diagonalization method in quantum recursion theory}

\author{Karl Svozil}
\email{svozil@tuwien.ac.at}
\homepage{http://tph.tuwien.ac.at/~svozil}
\affiliation{Institute for Theoretical Physics, Vienna University of Technology,  \\
Wiedner Hauptstra\ss e 8-10/136, A-1040 Vienna, Austria}

\begin{abstract}
As quantum parallelism allows the effective co-representation of classical mutually exclusive states, the diagonalization method of classical recursion theory has to be modified. Quantum diagonalization involves unitary operators whose eigenvalues are different from one.
\end{abstract}


\maketitle

\section{Introduction}

The reasoning in formal logic and the theory of recursive functions and
effective computability~\cite{rogers1,davis,Barwise-handbook-logic,enderton72,odi:89,Boolos-07},
at least insofar as their applicability to worldly things is concerned~\cite{landauer},
makes implicit assumptions about the physical meaningfulness  of the entities of
discourse; e.g., their actual physical representability and operationalizability~\cite{bridgman}.
It is this isomorphism or correspondence between the phenomena and  theory and {\it vice versa}
--- postulated by the Church-Turing thesis~\cite{Olszewski-06} ---
which confers power to the formal methods.
Therefore, any finding in physics presents a challenge to the formal sciences; at least
insofar as they claim to be relevant to the physical universe,
although history shows that the basic postulates have to be re-considered very rarely.

For example, the fundamental atom of classical information, the bit,
is usually assumed to be in one of two possible mutually exclusive states, which
can be represented by two distinct states of a classical physical system.
These issues have been extensively discussed in the context of energy dissipation associated
with certain logical operations and
universal (ir)reversible computation~\cite{fred-tof-82,maxwell-demon,feynman,feynman-computation}.

In general, all varieties of physical states, as well as their evolution and transformations, are relevant
for propositional logic as well as for a generalized theory of information.
Quantum logic~\cite{birkhoff-36}, partial algebras~\cite{kochen2,kochen3},
empirical logic~\cite{Foulis-Piron-Randall,Randall-Foulis} and continuous time computations~\cite{CIEChapter2007}
are endeavors in this direction.
These states need not necessarily be mapped into or bounded by classical information.
Likewise, physical transformations and manipulations available, for instance, in quantum information
and classical continuum theory, may
differ from the classical paper-and-pencil operations modeled by universal Turing machines.
Hence, the computational methods available as ``elementary operations''
have to be adapted to cope with the additional physical capabilities~\cite{deutsch}.

Indeed, in what follows it is argued that,
as quantum theory offers nonclassical states and operators available in quantum information theory,
several long-held assumptions on the character and transformation of classical information have to be adapted.
As a consequence, the formal techniques in manipulating information
in the theory of recursive functions and effective computability have to be revised.
Particular emphasis is given to undecidability and the diagonalization method.

\section{Quantum information theory}

As several fine presentations of quantum information and computation theory exist (cf. Refs.~\cite{Gruska,nielsen-book,Spiller-01,Brylinski-2002,Hayashi-06,Imai-Hayashi-06,arXiv:0802.4155,jaeger-2007,mermin-07}
for a few of them), there is no need of an extended exposition.
In what follows, we shall mainly follow Mermin's notation~\cite{mermin-02,mermin-07}.
For the representation of both a single classical and quantum bit,
suppose a two-dimensional Hilbert space. (For physical purposes
a linear vector space endowed with a scalar product will be sufficient.)
Let the superscript ``$T$'' indicate transposition,
and let $\vert 0\rangle \equiv (1,0)^T$ and $\vert 1\rangle \equiv (0,1)^T$ be the orthogonal vector representations of the
classical states associated with``falsity'' and ``truth,'' or ``0'' and ``1,'' respectively.

From the varieties of properties featured by quantum information,
one is of particular importance for quantum recursion theory:
the ability to co-represent classically distinct, contradictory states of information {\it via} the generalized quantum bit state
\begin{equation}
\vert \psi \rangle = \alpha_0 \vert 0\rangle + \alpha_1 \vert 1\rangle \equiv \left(
\begin{array}{c}
\alpha_0\\
\alpha_1
\end{array}
\right),
\end{equation}
with the normalization $\vert \alpha_0 \vert^2  +\vert \alpha_1 \vert^2 =1$.
This feature is also known as {\em quantum parallelism},
alluding to the fact that $n$ quantum bits can co-represent $2^n$ classical mutually exclusive states
$\left\{ \vert i_1 i_2 \cdots i_n\rangle \mid i_j\in \{0,1\}, \; j=1,\ldots ,n\right\}$ of $n$ classical bits.

As will be argued below, recursion theoretic diagonalization can be symbolized by the diagonalization or ``not'' operator
$\textsf{\textbf{X}}=
\left(
\begin{array}{cc}
0&1\\
1&0
\end{array}
\right)$, transforming $\vert 0\rangle$ into $\vert 1\rangle$, and {\it vice versa}.
The eigensystem of the diagonalization  operator  $\textsf{\textbf{X}}$
is given by the two 50:50 mixtures of $\vert 0\rangle $ and $\vert 1\rangle $
with the two eigenvalues $1$ and $-1$; i.e.,
\begin{equation}
\textsf{\textbf{X}}\frac{1}{\sqrt{2}}\left( \vert 0\rangle \pm \vert 1\rangle   \right)
= \pm \frac{1}{\sqrt{2}}\left( \vert 0\rangle \pm \vert 1\rangle   \right) = \pm \vert \psi_\pm \rangle.
\end{equation}
In particular,  the state   $\vert \psi_+ \rangle$ associated with the eigenvalue $+1$
is a {\em fixed point} of the operator $\textsf{\textbf{X}}$.

Note that, provided that $\vert \psi \rangle \not\in \{\vert 0 \rangle,\vert 1 \rangle \}$, a
quantum bit is not in a pure classical state. Therefore,
any practical determination of the quantum bit amounts to a measurement of the state ``along'' one context~\cite{svozil-2008-ql}
or base, such as the base ``spanned'' by $\{\vert 0 \rangle,\vert 1 \rangle \}$.
 Any such {\em single} measurement will be
indeterministic (provided  that the basis does not coincide with $\{\vert \psi_+ \rangle,\vert \psi_- \rangle\}$);
in particular, $\vert \langle \psi_\pm \vert 0\rangle \vert^2=\vert \langle \psi_\pm \vert 1\rangle \vert^2= 1/2$.
That is, if the fixed point state and the measurement context mismatch,
by Born's postulate~\cite{born-26-1,zeil-05_nature_ofQuantum},
the outcome of a {\em single} measurement occurs indeterministically, unpredictably and at random.
Hence, in terms of the quantum states $\vert 0 \rangle$ and $\vert 1 \rangle$ corresponding to the classical states,
the fixed point remains indeterminate.

In what follows it is argued that, due to the superposition principle,
the quantum recursion theoretic
diagonalization method has to be reformulated as a fixed point argument.
Application of the diagonal operator $\textsf{\textbf{X}}$ yields no {\em reductio ad absurdum.}
Instead, undecidability is recovered as a natural consequence of quantum coherence and of the
unpredictability of certain quantum events.

\section{Diagonalization}

For comprehensive reviews of recursion theory and the
diagonalization method the reader
is referred to Refs.~\cite{rogers1,davis,Barwise-handbook-logic,enderton72,odi:89,Boolos-07}.
Therefore, only a few hallmarks will be stated.
As already  pointed out by G\"odel
in his classical paper on the incompleteness of arithmetic~\cite{godel1},
the undecidability theorems of formal logic \cite{davis}
are based on semantical paradoxes
such as the liar
\cite{martin} or Richard's paradox.
A proper translation of the semantic paradoxes into formal proofs results in the diagonalization method.
Diagonalization has apparently first been applied by
Cantor to demonstrate the undenumerability of real numbers~\cite{cantor}. It has also been used by Turing for a proof
of the recursive undecidability of the halting problem~\cite{turing-36}.

A brief review of the classical algorithmic argument will be given
first.  Consider a universal computer $C$. For the sake of
contradiction, consider an arbitrary algorithm
$B(X)$ whose input is a string of symbols $X$.  Assume that there exists
a ``halting algorithm'' ${\tt HALT}$ which is able to decide whether $B$
terminates on $X$ or not.
The domain of ${\tt HALT}$  is the set of legal programs.
The range of ${\tt HALT}$ are classical bits (classical case) and quantum bits (quantum
mechanical case).

Using ${\tt HALT}(B(X))$ we shall construct another deterministic
computing agent $A$, which has as input any effective program $B$ and
which proceeds as follows:  Upon reading the program $B$ as input, $A$
makes a copy of it.  This can be readily achieved, since the program $B$
is presented to $A$ in some encoded form
$\ulcorner B\urcorner $,
i.e., as a string of
symbols.  In the next step, the agent uses the code
$\ulcorner B\urcorner $
 as input
string for $B$ itself; i.e., $A$ forms  $B(\ulcorner B\urcorner )$,
henceforth denoted by
$B(B)$.  The agent now hands $B(B)$ over to its subroutine ${\tt HALT}$.
Then, $A$ proceeds as follows:  if ${\tt HALT}(B(B))$ decides that
$B(B)$ halts, then the agent $A$ does not halt; this can for instance be
realized by an infinite {\tt DO}-loop; if ${\tt HALT}(B(B))$ decides
that $B(B)$ does {\em not} halt, then $A$ halts.

The agent $A$ will now be confronted with the following paradoxical
task:  take the own code as input and proceed.

\subsection{Classical case}
 Assume that $A$ is
restricted to classical bits of information.
To be more specific,
assume that ${\tt HALT}$ outputs the code of a classical bit as follows
($\uparrow$ and $\downarrow$ stands for divergence and convergence,
respectively):
\begin{equation}
{\tt HALT} ( B(X) ) =\left\{
 \begin{array}{l}
\vert 0 \rangle \mbox{ if } B(X) \uparrow
\\
\vert 1 \rangle \mbox{ if } B(X) \downarrow \\
\end{array}
 \right.
\quad .
\label{el:halt}
\end{equation}

Then, whenever $A(A)$
halts, ${\tt HALT}(A(A))$ outputs $\vert 1\rangle $ and forces $A(A)$ not to halt.
Conversely,
whenever $A(A)$ does not halt, then ${\tt HALT}(A(A))$ outputs $\vert 0\rangle $
and steers
$A(A)$ into the halting mode.  In both cases one arrives at a complete
contradiction.  Classically, this contradiction can only be consistently
avoided by assuming the nonexistence of $A$ and, since the only
nontrivial feature of $A$ is the use of the peculiar halting algorithm
${\tt HALT}$, the impossibility of any such halting algorithm.

\subsection{Quantum mechanical case}
As has been argued above, in quantum information theory
a quantum bit may be in a linear coherent
superposition
of the two classical states $\vert 0\rangle$ and $\vert 1\rangle$.
Due to the superposition of classical bit
states, the usual {\it reductio ad absurdum} argument breaks down.
Instead, diagonalization procedures in
quantum information theory yield quantum bit solutions which are fixed points
of the associated unitary operators.

In what follows it will be demonstrated how the task of the agent $A$
can be performed consistently if
$A$ is allowed to process quantum information.
To be more specific, assume that the output of the hypothetical
``halting algorithm'' is a quantum bit
\begin{equation}
{\tt HALT} ( B(X) ) = \vert \psi \rangle
\quad .
\end{equation}
We may think of   ${\tt HALT} ( B(X) )$ as a universal computer $C'$
simulating $C$ and containing a dedicated {\em halting bit}, which it
the output of $C'$
at every (discrete) time cycle. Initially (at time zero),
this halting bit is prepared to be a 50:50 mixture of the
classical halting and non-halting states $\vert 0\rangle$ and $\vert 1\rangle$ with equal phase; i.e.,
$\vert \psi_+ \rangle$. If later $C'$ finds that $C$ converges
(diverges) on $B(X)$, then the halting bit of $C'$ is set to the
``classical'' values $\vert 0 \rangle$ or $\vert 1 \rangle$.

The emergence of fixed points can be demonstrated by a simple example.
Agent $A$'s diagonalization task can be formalized as
follows. Consider for the moment the action of diagonalization on the
classical bit states. (Since the quantum bit states are merely a linear coherent superposition
thereof, the action of diagonalization on quantum bits is straightforward.)
Diagonalization effectively transforms the classical bit value $\vert 0\rangle$ into $\vert 1\rangle$ and
{\it vice versa.}
Recall that in equation
(\ref{el:halt}),  the state
$\vert 1\rangle$ has been identified  with the halting state
and the state $\vert 0\rangle$ with the non-halting state.

 The evolution representing diagonalization (effectively, agent
$A$'s task) can be expressed by the unitary operator $D$ as
\begin{equation}
\textsf{\textbf{D}} \vert 0 \rangle  =  \vert 1 \rangle \mbox{ and }
\textsf{\textbf{D}} \vert 1 \rangle  =  \vert 0 \rangle\quad .
\end{equation}
Thus, $\textsf{\textbf{D}}$ acts essentially as a ${\tt not}$-gate corresponding to the operator $\textsf{\textbf{X}}$.
In the above state basis, $\textsf{\textbf{D}}$ can be represented as follows:
\begin{equation}
\textsf{\textbf{D}}=  \textsf{\textbf{X}}=
\left(
\begin{array}{cc}
0 & 1\\
1 & 0
\end{array}
\right) \quad .
\end{equation}
$\textsf{\textbf{D}}$ will be called {\em diagonalization} operator, despite the fact
that the only nonvanishing components are off-diagonal.

As has been pointed out earlier,
quantum information theory allows a linear coherent superposition
$\vert \psi \rangle $
of the
``classical'' bit states $\vert 0\rangle $ and $\vert 1 \rangle $.
$\textsf{\textbf{D}}$
has a
fixed point at the quantum bit state
\begin{equation}
\vert \psi_+ \rangle ={ {1\over \sqrt{2}}\left(\vert 0 \rangle +\vert 1 \rangle \right) }
\equiv
{1\over \sqrt{2}} \left(
\begin{array}{c}
1 \\
1
 \end{array}
\right) \quad .
\end{equation}
$\vert \psi_+ \rangle$
does not give rise to inconsistencies~\cite{svozil-paradox}:
If agent $A$ hands over the fixed point state
$\vert \psi_+ \rangle$ to the diagonalization
operator $\textsf{\textbf{D}}$, the same state
$\vert \psi_+ \rangle$ is recovered.
Stated differently, as long as the output of the ``halting
algorithm'' to input $A(A)$ is $\vert \psi_+ \rangle$, diagonalization does not
change it. Hence, even if the (classically) ``paradoxical'' construction
of diagonalization is maintained, quantum theory does not give rise to a
paradox, because the quantum range of solutions is larger than the
classical one.
Therefore,
standard proofs of the recursive unsolvability of the halting problem
do not apply if agent $A$ is allowed a quantum bit. The consequences for
quantum recursion theory are discussed below.

\section{Consequences for quantum recursion theory}

Several critical remarks are in order.
It should be noted that the fixed point quantum bit ``solution''
of the above halting problem is of not much practical help.
In particular, if one is interested in the ``classical'' answer whether
or not $A(A)$ halts,  then one ultimately has to perform an
irreversible measurement
on the fixed point state. This  causes a state reduction into the
classical states corresponding to $\vert 0 \rangle$ and $\vert 1 \rangle$.
Any single measurement will yield an indeterministic result.
There is a 50:50 chance that
the fixed point state will be either in $\vert 0 \rangle$ or $\vert 1 \rangle$, since as has been argued before,
$\vert \langle \psi_\pm \vert 0\rangle \vert^2=\vert \langle \psi_\pm \vert 1\rangle \vert^2= 1/2$.
Thereby, classical undecidability is recovered.

Thus, as far as problem solving is concerned, classical bits are not much of an
advance. If a classical information is required, then quantum bits are not
better than probabilistic knowledge. With regards to the question of
whether or not a computer halts, the ``solution''
is effectively equivalent to the throwing of a fair coin~\cite{diaconis:211}.
Therefore, the advance of quantum recursion theory over classical
recursion theory is not so much classical problem solving but {\em the
consistent representation of statements} which would give rise to
classical paradoxes.

The above argument used the continuity of quantum bit states as compared to the
two discrete classical bit states for a construction of fixed points of the
diagonalization operator. One could proceed a step further and allow
{\em nonclassical diagonalization procedures}. Thereby, one could extend diagonalization to
the entire range of two-dimensional unitary transformations~\cite{murnaghan},
which need not have fixed points corresponding to eigenvalues of exactly one.
Note that the general diagonal form of finite-dimensional unitary transformations
in matrix notation is $\text{diag}(e^{i\varphi_1},e^{i\varphi_2}, \ldots, e^{i\varphi_n})$;
i.e., the eigenvalues of a unitary operator are complex numbers of unit modulus (e.g., Ref.~\cite[p.~39]{b5.171}, or Ref.~\cite[p.~161]{halmos-vs}).
Fixed points only occur if at least one of the phases $\varphi_i$, $i\in \{1, 2,\ldots ,n\}$ is a multiple of $2\pi$.
In what follows, we shall study the physical realizability of general unitary operators associated with
generalized beam splitters~\cite{rzbb,reck-94,zukowski-97,svozil-2004-analog}.
We will be particularly interested in those transformations whose spectra do not contain the eigenvalue one
and thus do not allow a fixed point eigenvector.

\begin{figure}
\begin{center}
\unitlength=0.50mm
\linethickness{0.4pt}
\begin{picture}(120.00,200.00)
\put(20.00,120.00){\framebox(80.00,80.00)[cc]{}}
\put(57.67,160.00){\line(1,0){5.00}}
\put(64.33,160.00){\line(1,0){5.00}}
\put(50.67,160.00){\line(1,0){5.00}}
\put(78.67,170.00){\framebox(8.00,4.33)[cc]{}}
\put(82.67,178.00){\makebox(0,0)[cc]{$P_3,\varphi$}}
\put(73.33,160.00){\makebox(0,0)[lc]{$S(T)$}}
\put(8.33,185.67){\makebox(0,0)[cc]{$\vert 0\rangle$}}
\put(110.67,185.67){\makebox(0,0)[cc]{${\vert 0\rangle}'$}}
\put(110.67,145.67){\makebox(0,0)[cc]{${\vert 1\rangle}'$}}
\put(8.00,145.67){\makebox(0,0)[cc]{$\vert 1\rangle$}}
\put(24.33,195.67){\makebox(0,0)[lc]{${\textsf{\textbf{T}}}^{bs}(\omega ,\alpha ,\beta ,\varphi )$}}
\put(0.00,179.67){\vector(1,0){20.00}}
\put(0.00,140.00){\vector(1,0){20.00}}
\put(100.00,180.00){\vector(1,0){20.00}}
\put(100.00,140.00){\vector(1,0){20.00}}
\put(20.00,14.67){\framebox(80.00,80.00)[cc]{}}
\put(20.00,34.67){\line(1,1){40.00}}
\put(60.00,74.67){\line(1,-1){40.00}}
\put(20.00,74.67){\line(1,-1){40.00}}
\put(60.00,34.67){\line(1,1){40.00}}
\put(55.00,74.67){\line(1,0){10.00}}
\put(55.00,34.67){\line(1,0){10.00}}
\put(37.67,54.67){\line(1,0){5.00}}
\put(44.33,54.67){\line(1,0){5.00}}
\put(30.67,54.67){\line(1,0){5.00}}
\put(77.67,54.67){\line(1,0){5.00}}
\put(84.33,54.67){\line(1,0){5.00}}
\put(70.67,54.67){\line(1,0){5.00}}
\put(88.67,64.67){\framebox(8.00,4.33)[cc]{}}
\put(93.67,73.67){\makebox(0,0)[rc]{$P_4,\varphi$}}
\put(60.00,80.67){\makebox(0,0)[cc]{$M$}}
\put(59.67,29.67){\makebox(0,0)[cc]{$M$}}
\put(28.67,57.67){\makebox(0,0)[rc]{$S_1$}}
\put(88.33,57.67){\makebox(0,0)[lc]{$S_2$}}
\put(8.33,80.34){\makebox(0,0)[cc]{$\vert 0\rangle$}}
\put(110.67,80.34){\makebox(0,0)[cc]{${\vert 0\rangle}'$}}
\put(110.67,40.34){\makebox(0,0)[cc]{${\vert 1\rangle}'$}}
\put(8.00,40.34){\makebox(0,0)[cc]{$\vert 1\rangle$}}
\put(49.00,39.67){\makebox(0,0)[cc]{$c$}}
\put(71.33,68.67){\makebox(0,0)[cc]{$b$}}
\put(24.33,90.34){\makebox(0,0)[lc]{${\textsf{\textbf{T}}}^{MZ}(\alpha ,\beta ,\omega,\varphi )$}}
\put(0.00,74.34){\vector(1,0){20.00}}
\put(0.00,34.67){\vector(1,0){20.00}}
\put(100.00,74.67){\vector(1,0){20.00}}
\put(100.00,34.67){\vector(1,0){20.00}}
\put(48.67,64.67){\framebox(8.00,4.33)[cc]{}}
\put(56.67,60.67){\makebox(0,0)[lc]{$P_3,\omega$}}
\put(10.00,110.00){\makebox(0,0)[cc]{a)}}
\put(10.00,4.67){\makebox(0,0)[cc]{b)}}
\put(20.00,140.00){\line(2,1){80.00}}
\put(20.00,180.00){\line(2,-1){80.00}}
\put(32.67,170.00){\framebox(8.00,4.33)[cc]{}}
\put(36.67,182.00){\makebox(0,0)[cc]{$P_1,\alpha +\beta $}}
\put(24.67,64.67){\framebox(8.00,4.33)[cc]{}}
\put(24.67,73.67){\makebox(0,0)[lc]{$P_1,\alpha +\beta$}}
\put(24.67,41.67){\framebox(8.00,4.33)[cc]{}}
\put(31.34,35.67){\makebox(0,0)[cc]{$P_2,\beta$}}
\put(32.67,147.00){\framebox(8.00,4.33)[cc]{}}
\put(36.67,155.00){\makebox(0,0)[cc]{$P_2,\beta$}}
\end{picture}
\end{center}
\caption{A universal quantum interference device operating on a qubit can be realized by a
4-port interferometer with two input ports ${\vert 0\rangle} ,{\vert 1\rangle} $
and two
output ports
${\vert 0\rangle} ',{\vert 1\rangle} '$;
a) realization
by a single beam
splitter $S(T)$
with variable transmission $T$
and three phase shifters $P_1,P_2,P_3$;
b) realization by two 50:50 beam
splitters $S_1$ and $S_2$ and four phase
shifters
$P_1,P_2,P_3,P_4$.
 \label{f:qid}}
\end{figure}
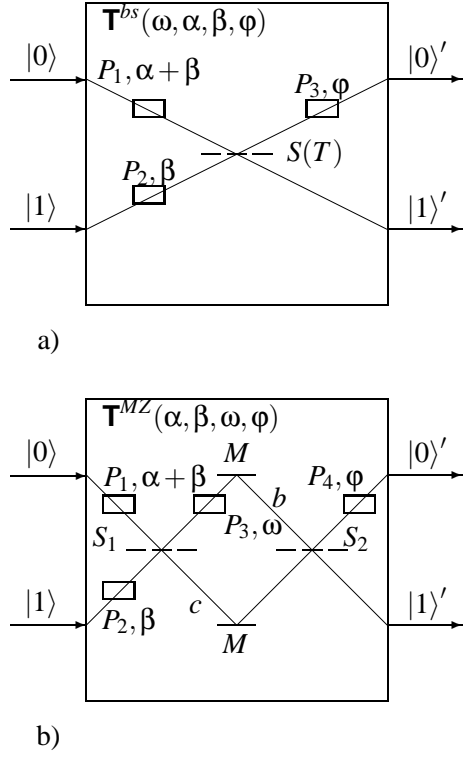

In what follows, lossless devices will be considered.
In order to be able to realize a universal unitary transformation in two-dimensional Hilbert space,
one needs to consider gates with two input und two output ports representing
beam splitters and  Mach-Zehnder interferometers equipped with an appropriate number of phase shifters.
For the sake of demonstration, consider the two realizations depicted in Fig.~\ref{f:qid}.
The
elementary quantum interference device ${\textsf{\textbf{T}}}^{bs}$  in
Fig.~\ref{f:qid}a)
is a unit consisting of two phase shifters $P_1$ and $P_2$ in the input ports, followed by a
beam splitter $S$, which is followed by a phase shifter  $P_3$ in one of the output
ports.
The device can
be quantum mechanically represented by \cite{green-horn-zei}
\begin{equation}
\begin{array}{rlcl}
P_1:&\vert {0}\rangle  &\rightarrow& \vert {0}\rangle e^{i(\alpha +\beta)}
 , \\
P_2:&\vert {1}\rangle  &\rightarrow& \vert {1}\rangle
e^{i \beta}
, \\
S:&\vert {0} \rangle
&\rightarrow& \sqrt{T}\,\vert {1}'\rangle  +i\sqrt{R}\,\vert {0}'\rangle
, \\
S:&\vert {1}\rangle  &\rightarrow& \sqrt{T}\,\vert {0}'\rangle  +i\sqrt{R}\,\vert
{1}'\rangle
, \\
P_3:&\vert {0}'\rangle  &\rightarrow& \vert {0}'\rangle e^{i
\varphi
} ,
\end{array}
\end{equation}
where
every reflection by a beam splitter $S$ contributes a phase $\pi /2$
and thus a factor of $e^{i\pi /2}=i$ to the state evolution.
Transmitted beams remain unchanged; i.e., there are no phase changes.
Global phase shifts from mirror reflections are omitted.
With
$\sqrt{T(\omega )}=\cos \omega$
and
$\sqrt{R(\omega )}=\sin \omega$,
the corresponding unitary evolution matrix
is given by
\begin{equation}
{\textsf{\textbf{T}}}^{bs} (\omega ,\alpha ,\beta ,\varphi )=
\left(
\begin{array}{cc}
 i \,e^{i \,\left( \alpha + \beta + \varphi \right) }\,\sin \omega &
   e^{i \,\left( \beta + \varphi \right) }\,\cos \omega
\cr
   e^{i \,\left( \alpha + \beta \right) }\, \cos \omega&
i \,e^{i \,\beta}\,\sin \omega
 \end{array}
\right)
.
\label{e:quid1}
\end{equation}
Alternatively, the action of a lossless beam splitter may be
described by the matrix
\footnote{
The standard labeling of the input and output ports are interchanged,
therefore sine and cosine are exchanged in the transition matrix.}
$$
\left(
\begin{array}{cc}
i \, \sqrt{R(\omega )}& \sqrt{T(\omega )}
\\
\sqrt{T(\omega )}&  i\, \sqrt{R(\omega )}
 \end{array}
\right)
=
\left(
\begin{array}{cc}
i \, \sin \omega  & \cos \omega
\\
\cos \omega&  i\, \sin \omega
 \end{array}
\right)
.
$$
A phase shifter in two-dimensional Hilbert space is represented by
either
${\rm  diag}\left(
e^{i\varphi },1
\right)
$
or
${\rm  diag}
\left(
1,e^{i\varphi }
\right)
$.
 The action of the entire device consisting of such elements is
calculated by multiplying the matrices in reverse order in which the
quanta pass these elements \cite{yurke-86,teich:90}; i.e.,
\begin{equation}
{\textsf{\textbf{T}}}^{bs} (\omega ,\alpha ,\beta ,\varphi )=
\left(
\begin{array}{cc}
e^{i\varphi}& 0\\
0& 1
\end{array}
\right)
\left(
\begin{array}{cc}
i \, \sin \omega  & \cos \omega
\\
\cos \omega&  i\, \sin \omega
\end{array}
\right)
\left(
\begin{array}{cc}
e^{i(\alpha + \beta)}& 0\\
0& 1
\end{array}
\right)
\left(
\begin{array}{cc}
1&0\\
0& e^{i\beta }
\end{array}
\right).
\end{equation}

The
elementary quantum interference device ${\textsf{\textbf{T}}}^{MZ}$ depicted in
Fig.~\ref{f:qid}b)
is a Mach-Zehnder interferometer with {\em two}
input and output ports and three phase shifters.
The process can
be quantum mechanically described by
\begin{equation}
\begin{array}{rlcl}
P_1:&\vert {0}\rangle  &\rightarrow& \vert {0}\rangle e^{i
(\alpha +\beta )} , \\
P_2:&\vert {1}\rangle  &\rightarrow& \vert {1}\rangle e^{i
\beta} , \\
S_1:&\vert {1}\rangle  &\rightarrow& (\vert b\rangle  +i\,\vert
c\rangle )/\sqrt{2} , \\
S_1:&\vert {0}\rangle  &\rightarrow& (\vert c\rangle  +i\,\vert
b\rangle )/\sqrt{2}, \\
P_3:&\vert b\rangle  &\rightarrow& \vert b\rangle e^{i \omega },\\
S_2:&\vert b\rangle  &\rightarrow& (\vert {1}'\rangle  + i\, \vert
{0}'\rangle )/\sqrt{2} ,\\
S_2:&\vert c\rangle  &\rightarrow& (\vert {0}'\rangle  + i\, \vert
{1}'\rangle )/\sqrt{2} ,\\
P_4:&\vert {0}'\rangle  &\rightarrow& \vert {0}'\rangle e^{i
\varphi
}.
\end{array}
\end{equation}
The corresponding unitary evolution matrix
is given by
\begin{equation}
{\textsf{\textbf{T}}}^{MZ} (\alpha ,\beta ,\omega ,\varphi )=
i \, e^{i(\beta +{\omega \over 2})}\;\left(
\begin{array}{cc}
-e^{i(\alpha +  \varphi )}\sin {\omega \over 2}
&
e^{i  \varphi }\cos {\omega \over 2} \\
e^{i  \alpha }\cos {\omega \over 2}
&
\sin {{\omega }\over 2}
 \end{array}
\right)
.
\label{e:quid2}
\end{equation}
Alternatively, ${\textsf{\textbf{T}}}^{MZ}$ can be computed by matrix multiplication; i.e.,
\begin{equation}
\begin{array}{l}
{\textsf{\textbf{T}}}^{MZ} (\alpha ,\beta ,\omega ,\varphi )=  \\
\qquad
i \, e^{i(\beta +{\omega \over 2})}\;
\left(
\begin{array}{cc}
e^{i\varphi }& 0\\
0& 1
 \end{array}
\right)
{1\over \sqrt{2}}\left(
\begin{array}{cc}
i& 1\\
1& i
 \end{array}
\right)
\left(
\begin{array}{cc}
e^{i\omega}& 0\\
0&1
 \end{array}
\right)
{1\over \sqrt{2}}\left(
\begin{array}{cc}
i& 1\\
1& i
 \end{array}
\right)
\left(
\begin{array}{cc}
e^{i(\alpha+\beta )}& 0\\
0&1
 \end{array}
\right)
\left(
\begin{array}{cc}
1& 0\\
0& e^{i\beta}
 \end{array}
\right)
 .
 \end{array}
\label{e:quid2mm}
\end{equation}
Both elementary quantum interference devices
${\textsf{\textbf{T}}}^{bs}$
and
${\textsf{\textbf{T}}}^{MZ}$
are  universal in the
sense that
 every unitary quantum
evolution operator in two-dimensional Hilbert space
\begin{equation}
\textsf{\textbf{U}}_2(\omega ,\alpha ,\beta ,\varphi )=e^{-i\,\beta}\,
\left(
\begin{array}{cc}
{e^{i\,\alpha }}\,\cos \omega
&
{-e^{-i\,\varphi }}\,\sin \omega
\\
{e^{i\,\varphi }}\,\sin \omega
&
{e^{-i\,\alpha }}\,\cos \omega
 \end{array}
\right)
 \quad ,
\label{e:quid3}
\end{equation}
where $-\pi \le \beta ,\omega \le \pi$,
$-\, {\pi \over 2} \le  \alpha ,\varphi \le {\pi \over 2}$~\cite{murnaghan}
corresponds to
${\textsf{\textbf{T}}}^{bs}
\left(\omega' ,
\alpha',
\beta' ,
\varphi'
\right)$
and
${\textsf{\textbf{T}}}^{MZ}
\left(\omega'' ,
\alpha'' ,
\beta'' ,
\varphi''
\right)$, where $\omega ,\alpha ,\beta ,\varphi $ are arguments of the (double) primed parameters~\cite{svozil-2004-analog}.

A typical example of a nonclassical operation on a quantum bit is
the ``square root of not'' gate
($
\sqrt{{\tt not}}
\sqrt{{\tt not}} =\textsf{\textbf{X}}$)
\begin{equation}
\sqrt{{\tt not}} =
{1 \over 2}
\left(
\begin{array}{cc}
1+i&1-i
\\
1-i&1+i
 \end{array}
\right)
\quad .
\end{equation}
Although $
\sqrt{{\tt not}}
$
still has a eigenstate  associated with a fixed point of unit eigenvalue,
not all of these unitary transformations have eigenvectors
associated with eigenvalues one that can be identified with fixed points.
Indeed, only unitary transformations of the form
\begin{equation}
\begin{array}{l}
[\textsf{\textbf{U}}_2(\omega ,\alpha ,\beta ,\varphi )]^{-1}\,\mbox{diag}(1, e^{i\lambda
}) \textsf{\textbf{U}}_2(\omega ,\alpha ,\beta ,\varphi )=  \\
\qquad
\qquad
\qquad
\left(
\begin{array}{cc}
{{\cos \omega }^2} + {e^{i\,\lambda }}\,{{\sin \omega }^2}&
{{{
{-1 + {e^{i\,\lambda
}}\over 2}
e^{-i\,\left(\alpha +\varphi \right) }}\,
\, \sin (2\,\omega )}} \\
{ -1 + {e^{i\,\lambda }}\over 2}
 {{{e^{i\,\left(\alpha
+\varphi \right) }}\,
 \sin
(2\,\omega )}}&
{e^{i\,\lambda }}\,{{\cos \omega }^2} + {{\sin
\omega }^2}
 \end{array}
\right)
 \end{array}
\end{equation}
have fixed points.

Applying nonclassical operations on quantum bits with no fixed points
\begin{equation}
\begin{array}{l}
\textsf{\textbf{D}}^\ast =[\textsf{\textbf{U}}_2(\omega ,\alpha ,\beta ,\varphi )]^{-1}\,\mbox{diag}( e^{i\mu } ,
e^{i\lambda }) \textsf{\textbf{U}}_2(\omega ,\alpha ,\beta ,\varphi )=  \\
\qquad
\qquad
\qquad
\left(
\begin{array}{cc}
  {e^{i\,\mu }}\,{{\cos (\omega )}^2} +
     {e^{i\,\lambda }}\,{{\sin (\omega )}^2}&
    {{{e^{-i\,\left( \alpha  + p \right) }\over 2}}\,
         \left( {e^{i\,\lambda }} - {e^{i\,\mu }} \right) \,\sin
(2\,\omega )}
       \\
{{{e^{i\,\left( \alpha  + p \right) }\over 2}}\,
        \left( {e^{i\,\lambda }} - {e^{i\,\mu }}  \right) \,\sin
(2\,\omega )}
       &{e^{i\,\lambda }}\,{{\cos (\omega )}^2} +
     {e^{i\,\mu }}\,{{\sin (\omega )}^2}
 \end{array}
\right) ,
 \end{array}
\end{equation}
with $\mu ,\lambda \neq 2n\pi$, $n\in {\Bbb N}_0$ gives rise to
eigenvectors which are not fixed points, but which acquire nonvanishing
phases $\mu , \lambda$ in the generalized diagonalization process.

\section{Summary}

It has been argued that, because of quantum parallelism,
i.e., the effective co-representation of classical mutually exclusive states, the diagonalization method of classical recursion theory has to be modified.
Quantum diagonalization involves unitary operators whose eigenvalues carry phases strictly different from multiples of $2 \pi$.
The quantum fixed point ``solutions'' of halting problems can be 50:50 mixtures of the classical halting
and nonhalting states, and therefore do not contribute to classical deterministic solutions of the associated decision problems.

Another, less abstract, application for quantum information theory is
the handling of inconsistent information in databases.
Thereby,
two contradicting classical bits of information
$\vert 0 \rangle$ and
$\vert 1 \rangle$ are resolved, i.e., co-represented, by the quantum bit
$\vert \psi_+ \rangle$.
Throughout the rest of the computation the coherence is maintained.
After the processing, the result is obtained by an irreversible
measurement. The processing of quantum bits, however, would require an
exponential
space overhead on classical computers in classical bit base \cite{feynman}.
Thus, in order to remain tractable,
the corresponding quantum bits should be implemented on
truly quantum universal computers.


\begin{thebibliography}{10}
\newcommand{\enquote}[1]{``#1''}
\expandafter\ifx\csname url\endcsname\relax
  \def\url#1{{#1}}\fi
\expandafter\ifx\csname urlprefix\endcsname\relax\def\urlprefix{}\fi

\bibitem{rogers1}
H.~{Rogers, Jr.}, {\em Theory of Recursive Functions and Effective
  Computability\/} (MacGraw-Hill, New York, 1967).

\bibitem{davis}
M.~Davis, {\em The Undecidable. Basic Papers on Undecidable, Unsolvable
  Problems and Computable Functions\/} (Raven Press, Hewlett, N.Y., 1965).

\bibitem{Barwise-handbook-logic}
J.~Barwise, {\em Handbook of Mathematical Logic\/} (North-Holland, Amsterdam,
  1978).

\bibitem{enderton72}
H.~Enderton, {\em {A Mathematical Introduction to Logic}\/} ({Academic Press},
  San Diego, 2001), second edn.

\bibitem{odi:89}
P.~Odifreddi, {\em Classical Recursion Theory, Vol. 1\/} (North-Holland,
  Amsterdam, 1989).

\bibitem{Boolos-07}
G.~S. Boolos, J.~P. Burgess, and R.~C. Jeffrey, {\em Computability and Logic\/}
  (Cambridge University Press, Cambridge, 2007), fifth edn.

\bibitem{landauer}
R.~Landauer, \enquote{Information is Physical,} Physics Today {\bf 44}, 23--29
  (1991).
\newline http://dx.doi.org/10.1063/1.881299

\bibitem{bridgman}
P.~W. Bridgman, \enquote{A Physicist's Second Reaction to {M}engenlehre,}
  Scripta Mathematica {\bf 2}, 101--117, 224--234 (1934), cf. R. Landauer
  \cite{landauer-95}.

\bibitem{Olszewski-06}
A.~Olszewski, J.~Wole{\'{n}}ski, and R.~Janusz, {\em {C}hurch's Thesis After 70
  Years\/} (Ontos, Berlin, 2006).

\bibitem{feynman}
R.~P. Feynman, \enquote{Simulating physics with computers,} International
  Journal of Theoretical Physics {\bf 21}, 467--488 (1982).

\bibitem{fred-tof-82}
E.~Fredkin and T.~Toffoli, \enquote{Conservative Logic,} International Journal
  of Theoretical Physics {\bf 21}, 219--253 (1982), reprinted in \cite[Part I,
  Chapter 3]{adama02}.
\newline http://dx.doi.org/10.1007/BF01857727

\bibitem{maxwell-demon}
H.~S. Leff and A.~F. Rex, {\em Maxwell's Demon\/} (Princeton University Press,
  Princeton, 1990).

\bibitem{feynman-computation}
R.~P. Feynman, {\em The Feynman lectures on computation\/} (Addison-Wesley
  Publishing Company, Reading, MA, 1996), edited by A.J.G. Hey and R. W. Allen.

\bibitem{birkhoff-36}
G.~Birkhoff and J.~von Neumann, \enquote{The Logic of Quantum Mechanics,}
  Annals of Mathematics {\bf 37}, 823--843 (1936).

\bibitem{kochen2}
S.~Kochen and E.~P. Specker, \enquote{Logical Structures arising in quantum
  theory,} in {\em Symposium on the Theory of Models, Proceedings of the 1963
  International Symposium at Berkeley\/}  pp. 177--189 (1965), reprinted in
  \cite[pp. 209--221]{specker-ges}.

\bibitem{kochen3}
S.~Kochen and E.~P. Specker, \enquote{The calculus of partial propositional
  functions,} in {\em Proceedings of the 1964 International Congress for Logic,
  Methodology and Philosophy of Science, Jerusalem\/}  pp. 45--57 (1965),
  reprinted in \cite[pp. 222--234]{specker-ges}.

\bibitem{Foulis-Piron-Randall}
D.~J. Foulis, C.~Piron, and C.~H. Randall, \enquote{Realism, operationalism,
  and quantum mechanics,} Foundations of Physics {\bf 13}, 813--841 (1983),
  invited papers dedicated to {G}{\"{u}}nther {L}udwig.
\newline http://dx.doi.org/10.1007/BF01906271

\bibitem{Randall-Foulis}
C.~H. Randall and D.~J. Foulis, \enquote{Properties and operational
  propositions in quantum mechanics,} Foundations of Physics {\bf 13}, 843--857
  (1983), invited papers dedicated to {G}{\"{u}}nther {L}udwig.
\newline http://dx.doi.org/10.1007/BF01906272

\bibitem{CIEChapter2007}
O.~Bournez and M.~L. Campagnolo, \enquote{A Survey on Continuous Time
  Computations,} in {\em New Computational Paradigms. Changing Conceptions of
  What is Computable\/}, S.~Cooper, B.~L{\"o}we, and A.~Sorbi, eds.  (Springer
  Verlag, New York, 2008), pp. 383--423.
\newline
  http://www.lix.polytechnique.fr/~bournez/pmwiki/uploads/Main/SurveyContinuou%
sTime.pdf

\bibitem{deutsch}
D.~Deutsch, \enquote{Quantum theory, the {C}hurch-{T}uring principle and the
  universal quantum computer,} Proceedings of the Royal Society of London.
  Series A, Mathematical and Physical Sciences (1934-1990) {\bf 400}, 97--117
  (1985).
\newline http://dx.doi.org/10.1098/rspa.1985.0070

\bibitem{Gruska}
J.~Gruska, {\em Quantum Computing\/} (McGraw-Hill, London, 1999).

\bibitem{nielsen-book}
M.~A. Nielsen and I.~L. Chuang, {\em Quantum Computation and Quantum
  Information\/} (Cambridge University Press, Cambridge, 2000).

\bibitem{Spiller-01}
H.-K. Lo, S.~Popescu, and T.~Spiller, {\em Introduction to Quantum Computation
  and Information\/} (World Scientific Publishing Company, Singapore, 2001).

\bibitem{Brylinski-2002}
R.~K. Brylinski, G.~Chen, and B.~K. Brylinski, {\em Mathematics of Quantum
  Computation\/} (Chapman \& Hall/CRC Press, London, 2002).

\bibitem{Hayashi-06}
M.~Hayashi, {\em Quantum Information. {A}n Introduction\/} (Springer-Verlag,
  Berlin, Heidelberg, 2006).

\bibitem{Imai-Hayashi-06}
H.~Imai and M.~Hayashi, {\em Quantum Computation and Information. {F}rom Theory
  to Experiment\/} (Springer-Verlag, Berlin, Heidelberg, 2006).

\bibitem{arXiv:0802.4155}
V.~Scarani, H.~Bechmann-Pasquinucci, N.~J. Cerf, M.~Dusek, N.~L{\"u}tkenhaus,
  and M.~Peev, \enquote{The Security of Practical Quantum Key Distribution,}
  (2008).
\newline http://arxiv.org/abs/0802.4155

\bibitem{jaeger-2007}
G.~Jaeger, {\em Quantum Information. {A}n Overview\/} (Springer, New York,
  2007).

\bibitem{mermin-07}
N.~D. Mermin, {\em Quantum Computer Science\/} (Cambridge University Press,
  Cambridge, 2007).
\newline http://people.ccmr.cornell.edu/~mermin/qcomp/CS483.html

\bibitem{mermin-02}
N.~D. Mermin, \enquote{From {C}bits to {Q}bits: Teaching computer scientists
  quantum mechanics,} American Journal of Physics {\bf 71}, 23--30 (2003).
\newline http://dx.doi.org/10.1119/1.1522741

\bibitem{svozil-2008-ql}
K.~Svozil, \enquote{Contexts in quantum, classical and partition logic,} in
  {\em Handbook of Quantum Logic and Quantum Structures\/}, K.~Engesser, D.~M.
  Gabbay, and D.~Lehmann, eds.  (Elsevier, Amsterdam, 2008), pp. 551--586.
\newline http://arxiv.org/abs/quant-ph/0609209

\bibitem{born-26-1}
M.~Born, \enquote{Zur {Q}uantenmechanik der {S}to{\ss}vorg{\"{a}}nge,}
  Zeitschrift f{\"{u}}r Physik {\bf 37}, 863--867 (1926).
\newline http://dx.doi.org/10.1007/BF01397477

\bibitem{zeil-05_nature_ofQuantum}
A.~Zeilinger, \enquote{The message of the quantum,} Nature {\bf 438}, 743
  (2005).
\newline http://dx.doi.org/10.1038/438743a

\bibitem{godel1}
K.~G{\"{o}}del, \enquote{{\"{U}}ber formal unentscheidbare {S\"{a}}tze der
  {P}rincipia {M}athematica und verwandter {S}ysteme,} Monatshefte f{\"{u}}r
  Mathematik und Physik {\bf 38}, 173--198 (1931), {E}nglish translation in
  \cite{godel-ges1}, and in \cite{davis}.

\bibitem{martin}
A.~R. Anderson, \enquote{St. {P}aul's epistle to {T}itus,} in {\em The Paradox
  of the Liar\/}, R.~L. Martin, ed.  (Yale University Press, New Haven, 1970),
  the Bible contains a passage which refers to Epimenides, a Crete living in
  the capital city of Cnossus: {\it ``One of themselves, a prophet of their
  own, said, `Cretans are always liars, evil beasts, lazy gluttons.'~''},---
  St. Paul, Epistle to Titus I (12-13).

\bibitem{cantor}
G.~Cantor, \enquote{{G}esammelte {A}bhandlungen,}  (Springer, Berlin, 1932).

\bibitem{turing-36}
A.~M. Turing, \enquote{On computable numbers, with an application to the
  {E}ntscheidungsproblem,} Proceedings of the London Mathematical Society,
  Series 2 {\bf 42 and 43}, 230--265 and 544--546 (1936-7 and 1937), reprinted
  in \cite{davis}.

\bibitem{svozil-paradox}
K.~Svozil, \enquote{Consistent use of paradoxes in deriving contraints on the
  dynamics of physical systems and of no-go-theorems,} Foundations of Physics
  Letters {\bf 8}, 523--535 (1995).

\bibitem{diaconis:211}
P.~Diaconis, S.~Holmes, and R.~Montgomery, \enquote{Dynamical Bias in the Coin
  Toss,} SIAM Review {\bf 49}, 211--235 (2007).
\newline http://dx.doi.org/10.1137/S0036144504446436

\bibitem{murnaghan}
F.~D. Murnaghan, {\em The Unitary and Rotation Groups\/} (Spartan Books,
  Washington, D.C., 1962).

\bibitem{b5.171}
R.~Shankar, {\em Principles of Quantum Mechanics, 2nd Edition\/} (Kluwer
  Academic/Plenum Publishers, New York, 1980,1994).

\bibitem{halmos-vs}
P.~R. Halmos, {\em Finite-dimensional vector spaces\/} (Springer, New York,
  Heidelberg, Berlin, 1974).

\bibitem{rzbb}
M.~Reck, A.~Zeilinger, H.~J. Bernstein, and P.~Bertani, \enquote{Experimental
  realization of any discrete unitary operator,} Physical Review Letters {\bf
  73}, 58--61 (1994).
\newline http://dx.doi.org/10.1103/PhysRevLett.73.58

\bibitem{reck-94}
M.~Reck and A.~Zeilinger, \enquote{Quantum phase tracing of correlated photons
  in optical multiports,} in {\em Quantum Interferometry\/}, F.~D. Martini,
  G.~Denardo, and A.~Zeilinger, eds.,  pp. 170--177 (1994).

\bibitem{zukowski-97}
M.~Zukowski, A.~Zeilinger, and M.~A. Horne, \enquote{Realizable
  higher-dimensional two-particle entanglements via multiport beam splitters,}
  Physical Review A (Atomic, Molecular, and Optical Physics) {\bf 55},
  2564--2579 (1997).
\newline http://dx.doi.org/10.1103/PhysRevA.55.2564

\bibitem{svozil-2004-analog}
K.~Svozil, \enquote{Noncontextuality in multipartite entanglement,} J. Phys. A:
  Math. Gen. {\bf 38}, 5781--5798 (2005).
\newline http://dx.doi.org/10.1088/0305-4470/38/25/013

\bibitem{green-horn-zei}
D.~M. Greenberger, M.~A. Horne, and A.~Zeilinger, \enquote{Multiparticle
  interferometry and the superposition principle,} Physics Today {\bf 46},
  22--29 (1993).

\bibitem{yurke-86}
B.~Yurke, S.~L. McCall, and J.~R. Klauder, \enquote{{SU(2)} and {SU(1,1)}
  interferometers,} Physical Review A (Atomic, Molecular, and Optical Physics)
  {\bf 33}, 4033--4054 (1986).
\newline http://dx.doi.org/10.1103/PhysRevA.33.4033

\bibitem{teich:90}
R.~A. Campos, B.~E.~A. Saleh, and M.~C. Teich, \enquote{Fourth-order
  interference of joint single-photon wave packets in lossless optical
  systems,} Physical Review A (Atomic, Molecular, and Optical Physics) {\bf
  42}, 4127--4137 (1990).
\newline http://dx.doi.org/10.1103/PhysRevA.42.4127

\bibitem{landauer-95}
R.~Landauer, \enquote{Advertisement For a Paper {I} Like,} in {\em On
  Limits\/}, J.~L. Casti and J.~F. Traub, eds.  (Santa Fe Institute Report
  94-10-056, Santa Fe, NM, 1994), p.~39.
\newline
  http://www.santafe.edu/research/publications/workingpapers/94-10-056.pdf

\bibitem{adama02}
A.~Adamatzky, {\em Collision-based computing\/} (Springer, London, 2002).

\bibitem{specker-ges}
E.~Specker, {\em Selecta\/} (Birkh{\"{a}}user Verlag, Basel, 1990).

\bibitem{godel-ges1}
K.~G{\"{o}}del, in {\em Collected Works. Publications 1929-1936. Volume {I}\/},
  S.~Feferman, J.~W. Dawson, S.~C. Kleene, G.~H. Moore, R.~M. Solovay, and
  J.~van Heijenoort, eds.  (Oxford University Press, Oxford, 1986).

\end{thebibliography}

\end{document}